\documentstyle{mn}
\newif\ifAMStwofonts
\ifoldfss
  \ifCUPmtlplainloaded \else
    \NewTextAlphabet{textbfit} {cmbxti10} {}
    \NewTextAlphabet{textbfss} {cmssbx10} {}
    \NewMathAlphabet{mathbfit} {cmbxti10} {} 
    \NewMathAlphabet{mathbfss} {cmssbx10} {} 
  \fi
  \ifAMStwofonts
    \ifCUPmtlplainloaded \else
      \NewSymbolFont{upmath} {eurm10}
      \NewSymbolFont{AMSa} {msam10}
      \NewMathSymbol{\upi}     {0}{upmath}{19}
      \NewMathSymbol{\umu}     {0}{upmath}{16}
      \NewMathSymbol{\upartial}{0}{upmath}{40}
      \NewMathSymbol{\leqslant}{3}{AMSa}{36}
      \NewMathSymbol{\geqslant}{3}{AMSa}{3E}

    \fi
  \fi
\fi 

\ifnfssone
  \newmathalphabet{\mathit}
  \addtoversion{normal}{\mathit}{cmr}{m}{it}
  \addtoversion{bold}{\mathit}{cmr}{bx}{it}
  \newmathalphabet{\mathbfit} 
  \addtoversion{normal}{\mathbfit}{cmr}{bx}{it}
  \addtoversion{bold}{\mathbfit}{cmr}{bx}{it}
  \newmathalphabet{\mathbfss} 
  \addtoversion{normal}{\mathbfss}{cmss}{bx}{n}
  \addtoversion{bold}{\mathbfss}{cmss}{bx}{n}
  \ifAMStwofonts
    \ifCUPmtlplainloaded \else
      \UseAMStwoboldmath
      \makeatletter
      \new@mathgroup\upmath@group
      \define@mathgroup\mv@normal\upmath@group{eur}{m}{n}
      \define@mathgroup\mv@bold\upmath@group{eur}{b}{n}
      \edef\UPM{\hexnumber\upmath@group}
      \new@mathgroup\amsa@group
      \define@mathgroup\mv@normal\amsa@group{msa}{m}{n}
      \define@mathgroup\mv@bold\amsa@group{msa}{m}{n}
      \edef\AMSa{\hexnumber\amsa@group}
      \makeatother
      \mathchardef\upi="0\UPM19
      \mathchardef\umu="0\UPM16
      \mathchardef\upartial="0\UPM40
      \mathchardef\leqslant="3\AMSa36
      \mathchardef\geqslant="3\AMSa3E
    \fi
  \fi
\fi 

\ifnfsstwo
  \DeclareMathAlphabet{\mathbfit}{OT1}{cmr}{bx}{it}
  \SetMathAlphabet\mathbfit{bold}{OT1}{cmr}{bx}{it}
  \DeclareMathAlphabet{\mathbfss}{OT1}{cmss}{bx}{n}
  \SetMathAlphabet\mathbfss{bold}{OT1}{cmss}{bx}{n}
  \ifAMStwofonts
    \ifCUPmtlplainloaded \else
      \DeclareSymbolFont{UPM}{U}{eur}{m}{n}
      \SetSymbolFont{UPM}{bold}{U}{eur}{b}{n}
      \DeclareSymbolFont{AMSa}{U}{msa}{m}{n}
      \DeclareMathSymbol{\upi}{0}{UPM}{"19}
      \DeclareMathSymbol{\umu}{0}{UPM}{"16}
      \DeclareMathSymbol{\upartial}{0}{UPM}{"40}
      \DeclareMathSymbol{\leqslant}{3}{AMSa}{"36}
      \DeclareMathSymbol{\geqslant}{3}{AMSa}{"3E}
    \fi
  \fi
\fi 

\ifCUPmtlplainloaded \else
  \ifAMStwofonts \else 
    \def\upi{\pi}
    \def\umu{\mu}
    \def\upartial{\partial}
  \fi
\fi

\pagerange{\pageref{firstpage}--\pageref{lastpage}}
\pubyear{1999}

\begin{document}
\twocolumn
\title[Nature of microstructure]{Nature of microstructure in pulsar radio emission}
\author[Machabeli et al.]{G. Machabeli$^{1}$, D. Khechinashvili$^{1,2}$, G. Melikidze$^{1,2}$ and D. Shapakidze$^{1}$\\
$^{1}$ Abastumani Astrophysical Observatory, Al. Kazbegi Avenue 2a, 380060
Tbilisi, Georgia\\
$^{2}$ J. Kepler Astronomical Center, Pedagogical University, Lubuska 2, 65-265
Zielona G\'ora, Poland}
\date{Accepted Received in original form}
\maketitle

\begin{abstract}
We present a model for microstructure in pulsar radio emission. We propose that micropulses result from the alteration of the radio wave generation region by nearly transverse drift waves propagating across the pulsar magnetic field and encircling the bundle of the open magnetic field lines. It is demonstrated that such waves can modify significantly curvature of these dipolar field lines. This in turn affects strongly fulfillment of the resonance conditions necessary for the excitation of radio waves. The time-scale of micropulses is therefore determined by the wavelength of drift waves. Main features of the microstructure are naturally explained in the frame of this model.
\end{abstract}


\label{firstpage} 

\section{Introduction}

Intensity variations of pulsar radio emission have several different time-scales, starting from a few nanoseconds (nanostructure). Single pulses consist of subpulses (there are usually about $3\div 5$ of them in a single pulse window) which often undergo drift, i.e. transition in phase. The nature of subpulses and most of their various features (such as circular polarisation, drifting subpulses, mode switching, nullings and phase memory phenomenon) were explained by the Plasma model for pulsar emission developed in the series of papers (see, e.g. Kazbegi et al. 1987, 1991a, 1991b, 1991c, 1996). In this paper we present an explanation of the pulsar microstructure in the frame of the Plasma model.

Let us first summarise the main observational features of microstructure. Micropulses are ultrashort intensity variations within individual pulses with the following properties. (1)~Characteristic time-scale is $20\div 30$~$\mu$s~\footnote{Time-scale generally ranges from 1~$\mu $s to 1~ms (Cordes 1979).}, with an upper limit of the time width distribution $0.1\div 1$ ms (Boriakoff 1996). (2)~Individual micropulse widths are constant with frequency. (3)~Micropulse widths are about twice smaller than the distance between micropulses. (4)~The time separation (phase) of any micropulse with respect to the fiducial point is constant with frequency, i.e. it is simultaneous at all frequencies at source (let us note here that the position of subpulses typically varies with the frequency). (5)~In about 25 per cent of the cases micropulses are observed in quasi-periodic sequences (trains) which can be detected simultaneously at widely separated frequencies, e.g. at 430 MHz and 1.4 GHz. (6)~The typical life-time of a micropulse does not exceed one second. (7)~Modulation depth depends on the frequency as $\nu^{-0.5}$.

Lange et al. (1998) reported on observations of microstructure at 4.85 and 1.41 GHz for a few pulsars. It appeared that for all observed pulsars a large fraction of the single pulses (varying between 30 and 70 per cent) show microstructure. For the pulsars where also low frequency results are available (Popov et al. 1988), there is no significant difference between microstructure properties at high and low frequencies. These results confirmed that microstructure is a common property of the pulsar emission and not only an additional feature of a few strong pulsars. Thus, the microstructure represents perhaps one of the fundamental features of pulsar pulses. Each self-consistent theory of pulsar radio emission should be able to explain existence of this phenomenon.

Two general types of models were presented for pulsar microstructure, often regarded as beaming models and temporal models, respectively (Chian \& Kennel 1983). In beaming models the observer's line of sight sweeps across a non-uniform pulsar beam, which results in rapid intensity fluctuation of observed radiation (Benford 1977; Ferguson 1981). In the angular beaming model proposed by Cordes (1979) and Gil (1982, 1986) both subpulses and micropulses are generated by the same narrow-band emission mechanism. Here micropulses correspond to thin plasma columns flowing along dipolar magnetic field lines, and the width of micropulses is determined by relativistic beaming. The model explains frequency stability of micropulses, in contrary to subpulses, basing on the geometrical consideration. The temporal models assume that the pulsar radiation is modulated while propagating through the magnetospheric plasma (see, e.g. Harding \& Tademaru 1981; Chian \& Kennel 1983).

Basing on the similarity of the microstructure time-scales over the broad frequency range, Lange et al. (1998) claimed that the micropulse duration is generated by the dimension of the emitting structure. Below we argue that the microstructure is caused by alteration of radio wave generation region by nearly transverse drift waves propagating across the magnetic field and encyrcling the open field lines region of the pulsar magnetosphere. Our mechanism explains naturally important features of microstructure.

\section{Emission model}
It is generally assumed that the pulsar magnetosphere is filled by dense relativistic electron-positron plasma flowing along the open magnetic field lines, which is generated as a consequence of the avalanche process first
described by Goldreich \& Julian (1969) and developed by Sturrock (1971). This plasma is multicomponent, with a one-dimensional distribution function (see Fig.~1 in Arons 1981), containing: (i) electrons and positrons of the
bulk of plasma with mean Lorentz factor of $\gamma _{p}$ and density $n_p$; (ii) particles of the high-energy 'tail' of the distribution function with $\gamma _{t}$ and $n_{t}$, stretched in the direction of positive momenta; (iii) the ultrarelativistic ($\gamma_{b}\sim 10^{6}$) primary beam with so called 'Goldreich-Julian' density $n_b \approx 7\times 10^{-2}B_{0}P^{-1}(R_{0}/r)^{3}~[{\rm cm}^{-3}]$ (where $P$ is a pulsar period, $R_{0}$ is a neutron star radius, $B_{0}$ is a magnetic field value at the stellar surface and $r$ is a distance from the neutron star's centre), which is much less than $n_{p}$ ($\kappa \equiv n_{p}/n_{b}\sim 10^{4\div 6}$). Such a distribution function should generate various wave-modes in certain conditions. These waves then propagate in the pair plasma of a pulsar magnetosphere, transform into the vacuum electromagnetic waves as the plasma density drops, enter the interstellar medium, and reach an observer as the pulsar radio emission.

An important feature of the pulsar radio emission is that the circular polarisation is observed in the vicinity of the maximum of a pulse (see, e.g. Taylor \& Stinebring 1986). At the same time, the exact maximum of the pulse power
corresponds to the zero circular polarisation (as the latter changes its sign). This means that the waves leave the magnetosphere propagating at relatively small angles to the pulsar magnetic field (Kazbegi et al. 1991b).

An extensive analysis have been conducted (Volokitin et al. 1985; Arons \& Barnard 1986; Lominadze et al. 1986) in order to study the dispersion properties of the waves propagating through the highly magnetised relativistic electron-positron plasma of pulsar magnetospheres. In the general case of oblique propagation with respect to the magnetic field three different wavemodes can be distinguished, these are: (i) the purely electromagnetic X-mode (also called $t$-wave), (ii) the subluminous Alfv\'{e}n mode ($lt$-wave) and iii) the superluminous O-mode ($L$-wave). The last two modes are of mixed electrostatic-electromagnetic nature. Electric field vectors ${\bmath E}^{{\rm O}}$ and ${\bmath E}^{{\rm A}}$ of the O and A-modes lie in the $\left( {\bmath k}\,{\bmath B} \right)$ plane, while the electric field of the X-mode ${\bmath E}^{{\rm X}}$ is directed perpendicularly to this plane. Here ${\bmath k}$ a wave-vector and ${\bmath B}$ is a local magnetic field.

Particles moving along the curved magnetic field undergo drift transversely to the plane of field curvature, with the velocity
\begin{equation}
u=\frac{c\gamma v_{\varphi }}{\omega_{_B}R_{c}},
\end{equation}
where $\omega_{_B}=(eB/mc)$, $R_{c}$ is the curvature radius of the dipolar magnetic field line, $\gamma$ is the relativistic Lorentz factor of a particle, and $v_{\varphi }$ is a particle velocity along the magneic field line. Here and below the cylindrical coordinate system $(x,~r,~\varphi )$ is chosed, with the $x$-axis directed transversely to the plane of the magnetic field curvature, while $r$ and $\varphi$ are the radial and azimuthal coordinates, respectively.

Generation of X and A-modes is possible after satisfaction of one of the following resonance conditions:
\begin{equation}
\omega - k_{\varphi }v_{\varphi } - k_{x}u=\frac{\omega_{_B}}{\gamma_{\mathrm res}},~ 
{\rm Cyclotron~instability,~and/or}
\end{equation}
\begin{equation}
\omega - k_{\varphi }v_{\varphi } - k_{x}u=0,~~~~
{\rm Cherenkov~instability}.
\end{equation}
These conditions are very sensitive to the parameters of the magnetospheric plasma, particularly to the value of the drift velocity (equation 1), hence to the curvature of magnetic field lines (Kazbegi et al. 1996).

Let us notice that the Plasma model is in accordance with observed systematic increase of component separation and profile widths with decreasing frequency, often called radius-to-frequency mapping (RFM). The RFM implies that the generation region is restricted to narrow range of altitudes. Besides, low-frequency waves are generated at relatively lower altitudes within this region. Namely, provided that magnetic field is dipolar far from the stellar surface, the wave frequency scales with the distance as
\begin{equation}
\omega _{0}\propto \left( \frac{R_{0}} {R}\right)^{6}
\end{equation}
in the case of the cyclotron (equation~2) mechanism, and as
\begin{equation}
\omega _{0}\propto \left( \frac{R_{0}}{R}\right)^{3.5}
\end{equation}
in the case of the Cherenkov (equation~3) mechanism (Kazbegi et al. 1991a). It is obvious that both mechanisms satisfy RFM, hence the emission region is localised to distinct altitudes.

\section{Generation of drift waves}
It was shown (Kazbegi et al. 1991c, 1996) that in addition to the X and A-modes (whose characteristic frequencies fall into radio band) propagating with small angles to the magnetic field lines, the very low-frequency nearly transverse drift waves can be excited. They propagate across the magnetic field, so that the angle $\theta$ between ${\bmath k}$ and ${\bmath B}$ is close to $\pi /2$. In other words, $k_{\perp}/k_{\varphi }\gg 1$, where $k_{\perp} = (k_{r}^{2}+k_{x}^{2})^{1/2}$. Assuming $\gamma (\omega /\omega_{_B})\ll 1,\ \left( u_{\alpha }/c\right) ^{2}\ll 1,\ k_{\varphi }/k_{x}\ll1\ $ and $k_{r}\rightarrow 0$, dispersion equation of the drift wave writes (Kazbegi et al. 1991c, 1996)
\begin{equation}
\frac{k_{x}^{2}c^{2}}{\omega ^{2}-k_{\varphi}^{2}c^{2}} = 1+\sum_{\alpha }\frac{\omega_{p\alpha }^{2}}
{\omega }\int \frac{v_{\varphi /c}}{\omega -k_{\varphi }v_{\varphi}-k_{x}u_{\alpha}}
\frac{\partial f}{\partial \gamma }d\gamma,
\end{equation}
where $\alpha$ denotes the sort of particles and $\omega _{p\alpha }^{2}=4\pi n_{p\alpha }e^{2}/m$.

Let us assume that 
\begin{equation}
\omega =k_{x}u_{b}+k_{\varphi }v_{\varphi }+a,
\end{equation}
where $u_{b}$ is a drift velocity of the beam particles (equation~1). Integration in parts and summation over the sorts of particles in equation (6), and use of equations~(7) yields
\begin{equation}
1-\frac{3}{2}\frac{1}{\gamma_{p}^{3}}\frac{\omega _{p}^{2}}{\omega ^{2}}-
\frac{1}{2}\frac{\omega _{p}^{2}}{\omega ^{3}}\frac{k_{x}u_{p}}{\gamma
_{p}}-\frac{\omega _{b}^{2}}{\omega a^{2}}\frac{k_{x}u_{b}}{\gamma _{b}}=
\frac{k_{\perp }^{2}c^{2}}{\omega ^{2}},
\end{equation}
where subscripts '$p$' and '$b$' denote values of the quantities corresponding to the bulk of plasma and the beam,
respectively. Note that a small term $(\omega _{b}^{2}/\gamma _{b}^{3}\omega a)$
has been neglected in equation (8). Thus, the frequency of a drift wave $\omega_0\equiv {\mathrm Re}\,\omega$ writes
\begin{equation}
\omega_0=k_{x}u_{b}+k_\varphi v_\varphi,
\end{equation}
where $k_\varphi v_\varphi \ll k_{x}u_{b}$.

Neglecting the second and the third terms in equation~(8), and solving this equation for the imaginary part of the complex frequency $|a|\approx\Gamma\equiv {\rm Im}\,\omega$, we obtain that the growth rate of drift waves is a maximum when
\begin{equation}
k_{\perp }^{2}\ll \frac{\omega _{p}^{2}}{\gamma _{p}^{3}c^{2}}.
\end{equation}
In this approximation the growth rate writes
\begin{equation}
\Gamma \simeq \left( \frac{n_{b}}{n_{p}}\right)^{1/2}
\left(\frac{\gamma_{p}^3}{\gamma_b}\right)^{1/2} k_{x}u_{b}.
\end{equation}

The drift waves propagate across the magnetic field and encircle the region of the open field lines of the pulsar magnetosphere. They draw energy from the longitudinal motion of the beam particles, as in the case of the
ordinary Cherenkov wave-particle interaction. However, they are excited only if $k_{x}u_{b}\neq 0$, i.e. in the presence of the drift motion of the beam particles. Note that these low-frequency waves are nearly transverse, with the electric vector directed almost along the local magnetic field.

\section{The model for microstructure}
Let us assume that a drift wave with the dispersion defined by equation (9) is excited at some place in the pulsar magnetosphere. It follows from the Maxwell equations that $B_{r}=E_{\varphi }\left(k_x c/\omega_0 \right)$, hence  $B_{r}\gg E_{\varphi}$ for such a wave. Therefore, excitation of a drift wave causes particular growth of the $r$-component of the local magnetic field.

The field line curvature $\rho _{c}\equiv 1/R_{c}$ is defined in a Cartesian frame of coordinates as 
\begin{equation}
\rho_{c}=\left[ 1+\left( \frac{dy}{dx}\right) ^{2}\right] ^{-3/2}\frac{d^{2}y}{dx^{2}},
\end{equation}
where $dy/dx=B_{y}/B_{x}$. Using $\left(\nabla \,{\bmath B}\right)=0$ and rewriting equation~(12) in the cylindrical coordinates we get 
\begin{equation}
\rho_{c}=\frac{1}{r}\frac{B_{\varphi }}{B}-\frac{1}{r}\frac{1}{B}\frac{B_{\varphi }^{2}}{B^{2}}\frac{\partial B_{r}}{\partial \varphi }.
\end{equation}
Here $B=\left( B_{\varphi }^{2}+B_{r}^{2}\right) ^{1/2}\approx B_{\varphi }
\left[ 1+(B_{r}^{2}/2B_{\varphi }^{2})\right] .$ Assuming that $k_{\varphi
}r\gg 1$ we obtain from equation (13)
\begin{equation}
\rho_{c}=\frac{1}{r}\left( 1 - k_{\varphi }r\frac{B_{r}}{B_{\varphi }}\right).
\end{equation}
From equation (14) it follows that even a small variation of $B_{r}$ causes significant change of $\rho_{c}$ (since $k_{\varphi}r\gg 1$).

Thus, the drift wave affects significantly the curvature of magnetic field lines, which in turn alters the drift velocity (equation~1) of particles. On the other hand, the resonance conditions~(2) and (3), which in the Plasma model are responsible for the radio wave generation in pulsar magnetospheres, are very sensitive to the parameters of the magnetospheric plasma, particularly to the value of the drift velocity $u$. Therefore, any variation of the magnetic field line curvature affects strongly the process of waves excitation. It follows that, within the generation region, the resonance conditions can be fulfilled only at the definite places ('emitting spots'), corresponding to the definite ('favorable') phases of the drift wave. The characteristic linear size $\lambda_{m}$ of a single emitting spot should be of order of the half-wavelength of the drift wave $\lambda_{m}\sim \lambda _{\perp}/2=\pi /k_{\perp}$. 

Assuming that the generation region corotates with a pulsar and introducing the altitude $R_{\mathrm em}$ of this region from the stellar surface, we can estimate the characteristic time during which the observer sweeps a single emitting spot
\begin{equation}
\tau _{m}=\frac{\lambda _{m}}{\Omega R_{\mathrm{em}}}\sim \frac{1}{2}\frac{P
}{k_{\perp }R_{\mathrm{em}}}\approx 10^{-10}\left( k_{\perp }{\mathcal R}_{\mathrm{em}}\right) ^{-1}\ [\mathrm{s}],
\end{equation}
where $\Omega =2\pi /P$ is the angular velocity of a pulsar and ${\mathcal R}_{\mathrm{em}}\equiv R_{\mathrm{em}}/R_{\mathrm{LC}}$ is a dimensionless altitude of the generation region from the stellar surface measured in the units of the light cylinder radius $R_{\mathrm{LC}}\approx 5\times 10^{9}P\ [\mathrm{cm}]$. Let us calculate the drift wave-number in the generation region of a 'typical' pulsar (i.e., the pulsar with $P=0.5$ s and $\dot{P}=10^{-15}$), for the magnetospheric parameters used in the Plasma model ($\gamma _{p}\approx 3$ and $\kappa \sim 10^{6}$). Note that according to this model radio emission originates at ${\mathcal R}_{\mathrm{em
}}=0.5\pm 0.2$. Thus, from equation (10) we obtain that
\begin{equation}
k_{\perp }\ll 3\times 10^{-3}\ \mathrm{{cm}}^{-1}.
\end{equation}
Speculating that $k_{\perp }\sim 10^{-6}\div 10^{-5}$ $\mathrm{{cm}}^{-1}$, in agreement with equation (16), we obtain from equation~(15) the characteristic times $\tau _{m}\sim 10^{-5}\div 10^{-4}$~s, which agree well with the typical time-scales of the microstructure. Note that the case $k_{\perp }\sim 10^{-5}~\mathrm{{cm}}^{-1}$ at ${\mathcal R}_{\mathrm{em}}\approx 0.5$ corresponds to the typical observed micropulse width of about $20~\mu\mathrm{s}$, whereas drift waves with the longer wavelengths $k_{\perp }\sim 10^{-6}~\mathrm{{cm}}^{-1}$
account for the time-scales corresponding to the upper limit of the microstructure widths distribution. From the model presented above naturally follow all the other features of the microstructure.

An observer distinguishes between the micropulses by their width which in turn is determined by the wave-length of the drift wave. On the other hand, the position of the emitting spots depends on the latitude, as it follows from equation (14). At the same time, the generation region, altered by the drift wave, is rather localised in both mechanisms (see equations 2 and 3), and the broad band of radio waves with different frequencies is excited there. So the observer detects different frequencies which are coming almost from the same place of the magnetosphere, i.e. from the same altitude and latitude. This explains the stability of the micropulse phase with respect to the pulse fiducial point at different frequencies. The quasi-periodic trains of micropulses observed at widely separated frequencies thus correspond to favorable phases of a drift wave.

From equation~(11) it follows that the growth rate of drift waves increases with the altitude from the stellar surface. Hence, the alteration of the generation region by these waves is stronger at higher altitudes where the lower frequencies are excited (equations~2 and 3). This explains why the probability of microstructure detection, as well as
its modulation depth, decrease at higher frequencies. 

Let us notice that although $k_{\varphi }\ll k_{x}$ for the drift waves, there still exists some non-zero $k_{\varphi }$. From equation~(9) it follows that the phase velocity of the drift wave has a longitudinal component
$v_{\varphi }^{\mathrm{ph}}/c\approx k_{\varphi }/k_{x}$ which gradually draws it away from the resonance region. This implies that the corresponding micropulse should disappear after a period of time 
\begin{equation}
\Delta t_{m}=\frac{\Delta R_{\mathrm{em}}}{v_{\varphi }^{\mathrm{ph}}}=\frac{
\Delta {\mathcal R}_{\mathrm{em}}}{2\pi }\frac{k_{x}}{k_{\varphi }}P,
\end{equation}
where $\Delta R_{\mathrm{em}}$ is a longitudinal size of the generation
region and $\Delta {\mathcal R}_{\mathrm{em}}$ is the same but in the units
of the light cylinder radius. Assuming that $\Delta {\mathcal R}_{\mathrm{em}}\approx 0.4$ we obtain for the 'typical' pulsar (with $P=0.5~$s) that $\Delta t_{m}\approx 1$~s if $k_{\varphi}/k_{x}\approx 0.03\ll 1$. This corresponds to the observed upper bound of the micropulse typical life-time.

\section{Summary}
In this paper we attempt to explain the existence and various properties of the microstructure in the pulsar radio emission basing on the Plasma model. According to suggested mechanism, the radio wave generation region is strongly altered by so called drift waves. These nearly transverse waves propagate across the local magnetic field, encyrcling the whole bundle of the open magnetic field lines, and cause significant change of field lines' curvature. This alters the curvature drift velocity (equation~1), therefore influencing fulfillment of the resonance conditions (equations~2 and 3) necessary for excitation of radio waves. Thus, radio waves are generated only at the definite regions (emitting spots) placed at approximately the same altitude from the star but different latitudes, corresponding to definite phases of the drift wave. The time-scale of the micropulses is thus determined by the characteristic time needed to the observer's line of sight to sweep an emitting spot with characteristic transverse size being about half-wavelength of the drift wave (equation 15). Characteristic times estimated from our model are in a good accordance with the typical observed time-scales of the microstructure.

Most of the important features of the microstructure are successfully explained by the model presented in this paper. However, we would like to stress that the results of the model are qualitative, and it does not pretend to strict quantitative explanation of all the existing observational data.

\section{Acknowledgments}
We thank T. Smirnova and V. Boriakoff for stimulating discussion. G. Ma. is thankful to J. Gil for his hospitality during the visit to J. Kepler Astronomical Centre, where the paper was completed. The work is supported in part by the INTAS Grant 96-0154. G. Me. and D. K. acknowledge also support from the KBN Grants 2 P03D 015 12 and 2 P03D 003 15. 

\label{lastpage}
\end{document}